# MC-UNet: Multi-module Concatenation based on U-shape Network for Retinal Blood Vessels Segmentation

Ting Zhang[a,b,†], Jun Li[a,b,†], Yi Zhao[a], Nan Chen[a,b], Han Zhou[a,b], Hongtao Xu[a,b], Zihao Guan[a,b], Changcai Yang[a,b]*, Lanyan Xue[a,b], Riqing Chen[a,b], Lifang Wei[a,b]*

[a] College of Computer and Information Science, Fujian Agriculture and Forestry University, Fuzhou 350002, China.

[b] Digital Fujian Research Institute of Big Data for Agriculture and Forestry, Fujian Agriculture and Forestry University, Fuzhou 350002, China.

† Co-first authors

* Corresponding author

*E-mail:* weilifang1981@163.com (Lifang Wei)

**Abstract:** Accurate segmentation of the blood vessels of the retina is an important step in clinical diagnosis of ophthalmic diseases. Many deep learning frameworks have come up for retinal blood vessels segmentation tasks. However, the complex vascular structure and uncertain pathological features make the blood vessel segmentation still very challenging. A novel U-shaped network named Multi-module Concatenation which is based on Atrous convolution and multi-kernel pooling is put forward to retinal vessels segmentation in this paper. The proposed network structure retains three layers the essential structure of U-Net, in which the atrous convolution combining the multi-kernel pooling blocks are designed to obtain more contextual information. The spatial attention module is concatenated with dense atrous convolution module and multi-kernel pooling module to form a multi-module concatenation. And different dilation rates are selected by cascading to acquire a larger receptive field in atrous convolution. Adequate comparative experiments are conducted on these public retinal datasets: DRIVE, STARE and CHASE_DB1. The results show that the proposed method is effective, especially for microvessels. The code will be put out at https://github.com/Rebeccala/MC-UNet

## 1. Introduction

The retinal is one of the most important parts of the eyes [1]. On the basis of the data published by the WHO, a growing number of people around the world are suffering from eye diseases [2]. The morphological characteristics of retinal blood vessels are very helpful for ophthalmologists who can use morphological features of retinal blood vessels, such as branching patterns, angles, curvatures, widths, and lengths, to diagnose and assess eye diseases [3-4]. Ophthalmologist can effectively screen and diagnose fundus related diseases by examining and analyzing the shape structure of retinal blood vessels. Therefore, fundus examination is an important part of ophthalmic examination. Extracting the shape and structure of retinal blood vessels is the most pivotal procedure in ophthalmic examination for ophthalmologists to identify diseases. In traditional medical procedures, the retinal vascular area needs to be manually segmented by experienced specialists that is time consuming and labor consuming. Moreover, the blood vessels in the retinal image are irregular and densely distributed, such as a lot of small blood vessels with low contrast, which is easily confused with the background. Although there are many retinal image segmentation methods which have been presented, those issues make the blood vessel segmentation still very challenging.

Unsupervised method and supervised learning method comprise retinal vessel segmentation method. The difference between them is whether the input data have manually segmented labels. Oliveira et al. [8] used two algorithms of median ranking and weighted mean which are different to combine Frangi filter, matched filter, and gabor wavelet filter for blood vessels segmentation. Alhussein et al. [9] extracted the enhanced images of thin and thick blood vessels respectively based on hessian matrix and intensity transformation method. Azzopardi et al. [10] presented a selective response vascular filter called B-COSFIRE for vascular segmentation. Saffarzadeh et al. [5] used multi-scale line operator to detect blood vessels, and used K-means to doblood vessels segmentation. These methods are efficient and fast in retinal vessel segmentation, but the segmentation performances are dependent on the selection of feature extractors. While supervised learning methods can learn features from the original images and segmentation labels that makes it more effective in segmentation tasks owing to get the input-output relationship. And the supervised learning methods can be subdivided into deep learning methods and traditional machine learning methods. The SVM and random forest which belong to traditional machine learning models need to manually construct features and map them to the target space. Wang [6] combined the characteristics of Gaussian scale space and the divergence characteristics of vector field, and used SVM classifier to segment blood vessels. Zhu et al. [7] used Cart and AdaBoost classifiers to classify pixels. Although the traditional machine learning method is easy to understand and can be explained, it requests to fit the feature types and feature selection methods that makes the feature representation ability is limited.

During the past few years, convolutional neural network (CNN) has made outstanding achievements in the automatic segmentation of retinal vessels. Compared with traditional machine learning, there are many layers of neural network in deep learning, which has strong nonlinear modeling ability and feature representation ability. In particular, since the U-Net [11] was proposed, various U-shaped networks based on encoding and decoding structure make biomedical images have more accurate segmentation performance. And several excellent retinal vessel segmentation methods which are U-NET based are proposed. Li et al. [12] proposed a method using structural redundancy in the vascular network to find fuzzy vascular details from the segmented vascular images and expand the depth of the model through multiple iterations. Two U-NET based models which is recurrent and recurrent residual have been proposed by Alom et al. [13], using the functions of residual network and RCNN. Zhuang et al. [14] proposes a multi-U-Net chain containing multiple encoder-decoder branches. Yuan et al. [21] fused the multi-level attention module with UNET to obtain the fusion information of low and high levels for alleviating the problem of network over fitting and obtaining generalization ability. Wang et al. [15] designed a dual-coding U-NET, which has outstanding performance in improving the segmentation capability of vessels in retina. A spatial attention module is added in the SA-UNET (Spatial Attention U-Net for Retinal Vessel Segmentation) to obtain more features of spatial dimensions by Guo et al. [16].

Although these U-Nets and their improved networks have been used in retinal vessels segmentation so widely, those suffers from many limitations and deficiencies. The encoder-decoder structures receive the information feature and its transmission in same layer by jump connections, which may cause the loss of small and fragile vessels owing to the limited comprehensive features. In order to alleviate the problems, we propose a multi-module concatenation network based on U-shape network called MC-UNet for retinal vessel segmentation, which retains local and global information about retinal main blood vessels and capillaries. Concretely, the contributions that this paper can make are summarized as follows:

(1) We raise a multi-kernel pooling based on the U-shape network that retains three layers the essential structure of U-Net but the atrous convolution combining the multi-kernel pooling blocks are designed to obtain more contextual information.

(2) We design a multi-module concatenation network to contain local and global information for retaining small vascular and advanced features.

(3) The spatial attention module in the network is concatenated with dense atrous convolution module and multi-kernel pooling module, which can further enhance the significance of the target.

(4) We evaluate and analyze the proposed MC-UNet on the challenging task of retinal blood vessels segmentation. According to the results of the experiments, our method reaches the state-of-the-art level on the public datasets.

## 2. Method

In this section, we will introduce our proposed MC-UNet shown in Fig.1. Our network retains three layers the essential structure of U-Net with a spatial attention module same as SA-UNET [16]. The Dropblock and BN modules are used to take the place of the convolution block in the original U-NET, which can effectively prevent the excessive merging of the network and improve network training speed. Consequently, it is more suitable for small sample data sets. And we bind the dense atrous convolution module (DAC) and multi-kernel pooling module (MKP), which joint local and global information for a certain extent. Then the spatial attention module in the network is concatenated with DAC and MKP. We will elaborate the MC-UNet in detail as the following subsections.

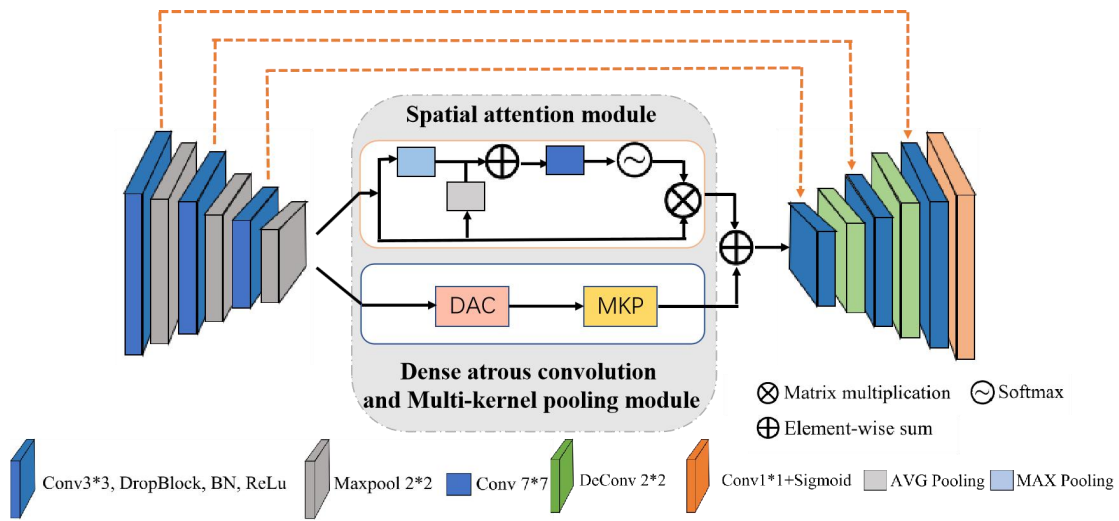

Fig.1 Diagram of the proposed MC-UNet

**Spatial attention module**. The spatial attention module [18] generates a spatial attention map using the maximum pool and average pool operations, selectively paying attention to the feature information in the image and ignoring other background information. The output feature $SA$ is obtained by multiplied of the input feature $F$ and attention map $\sigma(\cdot)$, which shown in formula (1). Where, $f^7$ and $\sigma$ represent 7*7 convolution operation and denote the Sigmoid Function, respectively. The illustrations of Spatial Attention module is shown in Fig.2.

$$SA = F \cdot \sigma(f^7([maxpool(F); Avgpool(F)])) \qquad (1)$$

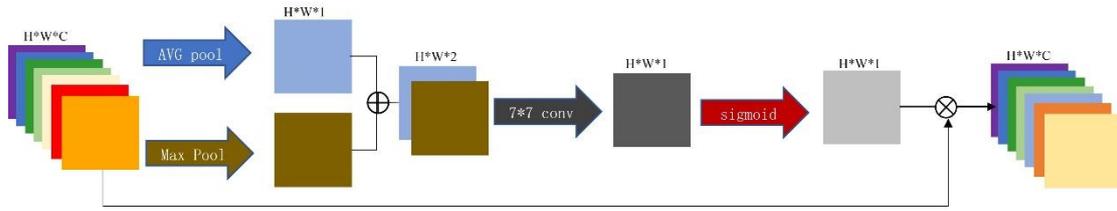

Fig. 2. The illustrations of Spatial Attention module (SA).

**Dense Atrous Convolution Module.** Atrous convolution has a widespread application in semantic segmentation, target detection and other tasks by many classical networks, such as Deeplab [19,20]. In deep learning algorithms, profit from pooling layer and convolution layer, the receptive field of feature image is increased and the size of feature image is reduced. What's more, upsampling is used to make the image size restored. But now, due to the process of feature image shrinkage and magnification, the accuracy will be lost. Atrous convolution can increase the receptive field and maintain the size of the feature map to reduce the computation of the network, which is utilized to replace down-sampling and up-sampling. The dilation rate of the atrous convolution can be set with different values, by which different receptive fields can be achieved for multi-scale information.

$$y(i) = \sum_k x[i + r * k]w[k] \qquad (2)$$

where $r$ represents the dilation rate. In particular, when $r = 1$, formula (2) is the standard convolution. The input feature map $x$ convolved with a filter $w$ obtain the output $y$. And Fig. 3 shows the schematic diagram of the atrous convolution, the dilation rates are 1, 3 and 5, respectively.

Compared with downsampling, atrous convolution can both enlarge the receptive field nicely and accurately locate the target and reduce the loss of spatial resolution. The dense atrous convolution [17] module shown in Fig. 4 is generated by integrating the atrous convolution using different dilation rates, which can capture the context information of different scales and achieve the local or global information. By using different dilation rates $r_k$ to combinate the atrous convolution, the output $D$ of atrous convolution modules can be obtained.

$$D = \sum_k y_{r_k}(x) \qquad (3)$$

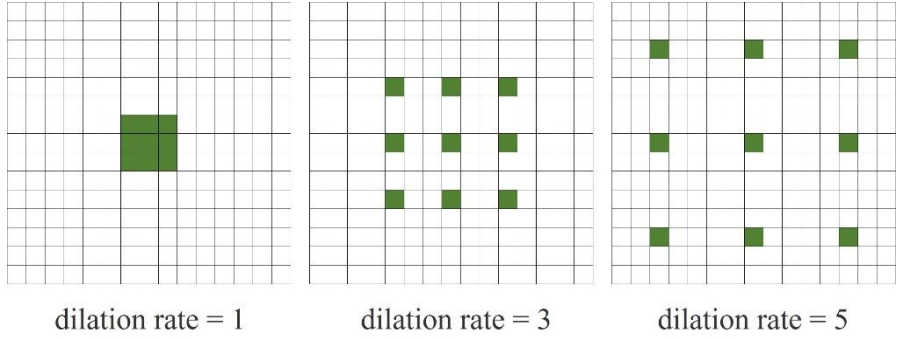

Fig. 3. Three atrous convolution of different dilation rate, the dilation rates are 1, 3 and 5, respectively.

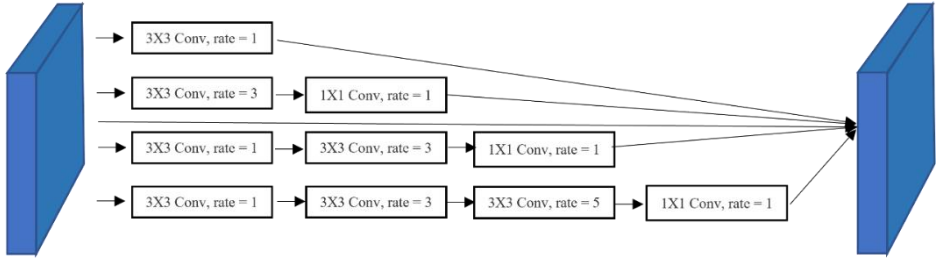

Fig. 4. The illustrations of dense atrous convolution (DAC)

**Multi-kernel pooling module.** Multi-kernel pooling [17] module is changed based on the spatial pyramid [25], which can make the redundant information of the feature map be redused and the amount of calculation. According to the different sizes of kernel, the feature information of receptive fields with different sizes is extracted to increase the segmentation performance of the model. The multi-kernel pooling module is introduced into the SA-UNET, which relies on multiple different kernels to detect different sizes targets. Multi-kernel pooling can use more context information by combining general max-pooling operation of different kernel size as shown in figure 5. And encoding the global context information into four receiving domains of different sizes: $2 \times 2, 3 \times 3, 5 \times 5$ and $6 \times 6$. Then, a 1 x 1 convolution is carried out to reduce the dimension of feature mapping, and upsampling is carried out to get features of the same size as the original feature mapping. Lastly, we concatenate the original features and the upsampled feature mapping and obtain the output feature $MKP$ of multi-kernel pooling module.

$$MKP = \sum_i f^1(Maxpool_{k_i}(D)) \qquad (4)$$

where, $f^1$ and $k_i$ denote the 1x1 convolution and $i$ kernels of different sizes, and the $D$ is the output feature map representing the dense atrous convolution module

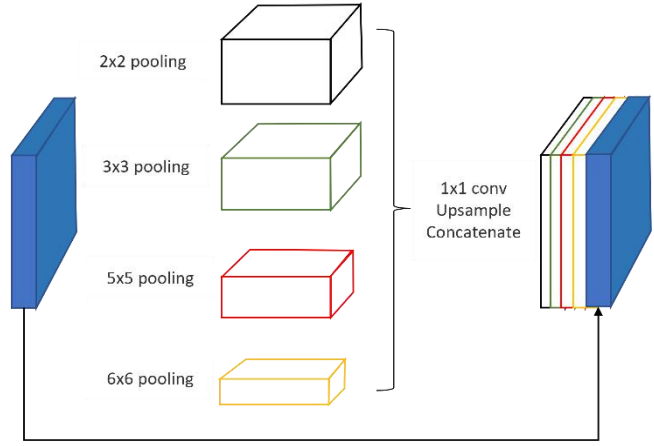

Fig. 5. The illustrations of residual multi-kernel pooling.

The encoder-decoder structures receive the information feature and its transmission in same layer by jump connections, which may cause the loss of small and fragile vessels owing to the limited comprehensive features. Spatial attention module, multi-kernel pooling module and dense atrous convolution module are complementary in the ability and scope of feature acquisition. Inspired by them, we propose a multi-module concatenation network for accurate retinal vessel segmentation. The output feature map $F$ is obtained by concatenating the output features of spatial attention module $SA$ and multi-kernel pooling module $MKP$.

$$F = SA + MKP \qquad (5)$$

## 3. Experiments

### 3.1 DataSets

We use the fundus data sets which are publicly available to verify our method, DRIVE [22] (digital retinal images for vessel extraction), CHASE_DB1 [24] (child heart and health study in England) and STARE [23] (structured analysis of the retina) to evaluate the segmentation performance of our approach MC-UNet. The STARE dataset includes pathological abnormal and healthy retinal images, which can evaluate the impact of the model on abnormal fundus images. The specific information of the three data sets is shown in Table 1.

Table 1. The details of the three datasets of DRIVE, CHASE_DB1 and STARE.

| DataSet | Resolution | Numbers of images | Train/Test Split |
|---|---|---|---|
| DRIVE | 565×584 | 40 | 20 images for training, 20 images for testing |
| CHASE_DB1 | 999×960 | 28 | 14 images for training, 14 images for testing |

| STARE | 700×605 | 20 | 10 images for training, 10 images for testing |

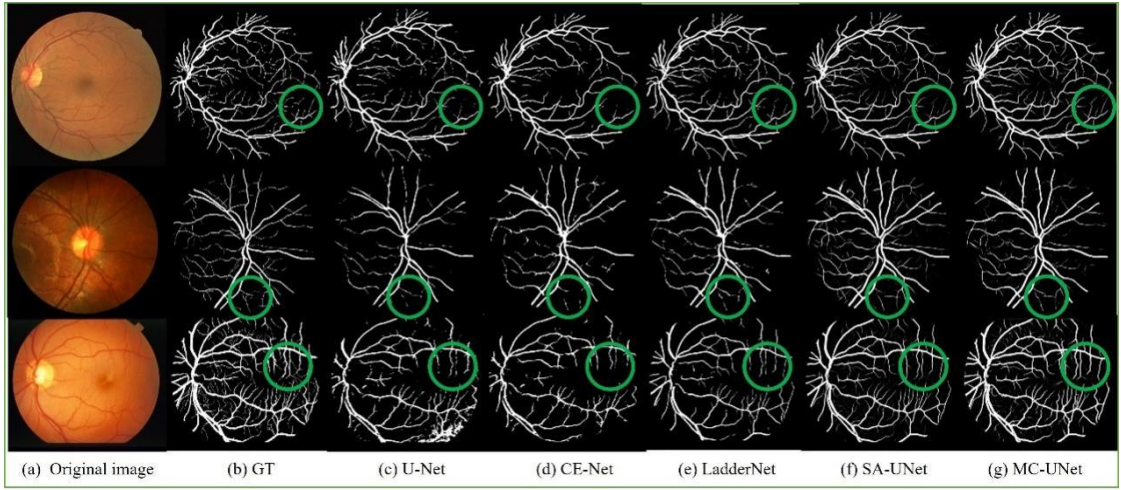

Fig 5. The segmentation example of the three datasets. Among them, (a) is the original retinal image, and (b) is the the ground truth. From (c) to (g) are segmentation maps by U-Net, CE-Net, LadderNet, SA-UNet and our method, respectively. The DRIVE dataset is shown in the first rowwhile the CHASEDB1 dataset is shown in the second row, and the STARE dataset is shown in the last row.

In the three datasets, we train and evaluate our method by using the manual annotation marked by the first expert. The segmentation result examples from DRIVE, STARE and CHASE_DB1 data sets are shown in Figure 5, which perceive the comparisons of the segmentation results on the three datasets with other methods are listed, including some methods based on U-Net. From Fig 5 (a) to Fig 5 (g), there are the original color retinal image, the ground truth, the segmentation map by U-Net [11], CE-Net [17], LadderNet [14], SA-Unet [16] and proposed method, respectively. Moreover, all the experiments were carried out on NVIDIA Quadro M5000 and 3.00 GHz PCs.

### 3.2 Evaluation Criteria

The aim of retinal vascular binary classification work is to divide each pixel in the input images into two categories: vascular (positive) and background (negative). By comparing the segmentation maps with the true value of label, four indexes can be obtained: TP, TN, FP and FN. P represents the number of white pixels in true images, and N represents the number of black pixels in the true image. T for true, F for false. TP represents the number of white pixels correctly predicted by optic disc, while TN represents the number of black pixels correctly predicted by optic disc. Refers to the total number of pixels in all truth images.

On the basis of these four basic indexes, accuracy (ACC), sensitivity (SEN), specificity (SP), area enclosed by coordinate axis (AUC) under ROC curve and F1-score can be calculated [16]. In our experiment, almost all the above indicators are used. The calculation formulas are as described below:

$$ACC = \frac{TP+TN}{TN+FP+TP+FN} \quad (3\text{-}1)$$

$$SE = \frac{TP}{TP+FN} \quad (3\text{-}2)$$

$$SP = \frac{TN}{FP+TN} \quad (3\text{-}3)$$

## 3.3 Results

We compare the segmentation result on the three datasets with other methods shown in Table 2. Notably, MC-UNet achieves the best performance on DRIVE and CHASE_DB1. And by comparing with the backbone our method has better performance, which illustrates that the proposed framework is effective for vascular segmentation. Specifically, the SE and AUC of our framework on three data sets are higher than backbone, our method has the highest ACC, SP and AUC on DRIVE, our method has the highest ACC, SE and AUC on CHASE_DB1. Due to many lesion images in the STARE dataset, the sensitivity index is not satisfactory on the STARE by MC-UNet. However, indicators are improved compared with the backbone network, which also verificates our method is effective.

Table 2. The Comparison of our model and other methods in DRIVE, STARE and CHASEDB1.

| DATASET | Method | Year | ACC | SEN | SP | AUC | F1 |
|---|---|---|---|---|---|---|---|
| DRIVE | U-net [11] | 2016 | 96.60 | 76.82 | 98.53 | 97.07 | - |
|  | R2u-net [13] | 2018 | 97.84 | 77.92 | 98.13 | 97.84 | - |
|  | Laddernet [14] | 2018 | 95.61 | 78.56 | 98.10 | 97.93 | 82.02 |
|  | Iternet [12] | 2019 | 95.74 | 77.91 | 98.17 | 98.16 | 82.18 |
|  | Ce-net [17] | 2019 | 95.50 | 79.03 | 97.69 | 97.80 | - |
|  | SA-Unet [16] | 2020 | 96.41 | **81.12** | 97.67 | 97.38 | 80.27 |
|  | AACA-MLA-D-UNet [21] | 2021 | 95.81 | 80.46 | 98.05 | 98.27 | **83.03** |
|  | MC-UNet | 2022 | **96.78** | 81.00 | **98.79** | **98.28** | 81.49 |
| STARE | U-net [11] | 2016 | 96.43 | 77.64 | 98.65 | 90.63 | - |
|  | R2u-net [13] | 2018 | 96.34 | 77.56 | 98.20 | 98.15 | - |
|  | Laddernet [14] | 2018 | 96.13 | 78.22 | 98.04 | 96.44 | 79.94 |
|  | Iternet [12] | 2019 | **97.60** | **79.69** | 98.23 | **98.37** | 80.73 |
|  | Ce-net [17] | 2019 | 97.32 | 79.09 | 97.21 | 95.97 | - |
|  | SA-Unet [16] | 2020 | 95.21 | 71.20 | 99.30 | 96.26 | 77.36 |
|  | AACA-MLA-D-UNet [21] | 2021 | 96.65 | 79.14 | 98.70 | 98.24 | **82.76** |
|  | MC-UNet | 2022 | 95.72 | 73.60 | **99.47** | 96.86 | 78.65 |
| CHASEDB1 | U-net [11] | 2016 | 96.43 | 77.64 | **98.65** | 93.26 | - |

| | | | | | |
|---|---|---|---|---|---|
| R2u-net [13] | 2018 | 96.34 | 77.56 | 98.20 | 98.15 | - |
| Laddernet [14] | 2018 | 96.56 | 79.78 | 98.18 | 96.94 | 80.31 |
| Iternet [11] | 2019 | 97.02 | 79.69 | 98.23 | 98.13 | 80.73 |
| Ce-net [17] | 2019 | 96.33 | 80.08 | 97.23 | 97.97 | - |
| SA-Unet [16] | 2020 | 97.08 | 81.51 | 98.09 | 97.78 | 77.36 |
| AACA-MLA-D-UNet [21] | 2021 | 96.73 | 83.02 | 98.01 | 98.10 | **82.48** |
| MC-UNet | 2022 | **97.14** | **83.66** | 98.29 | **98.18** | 77.41 |

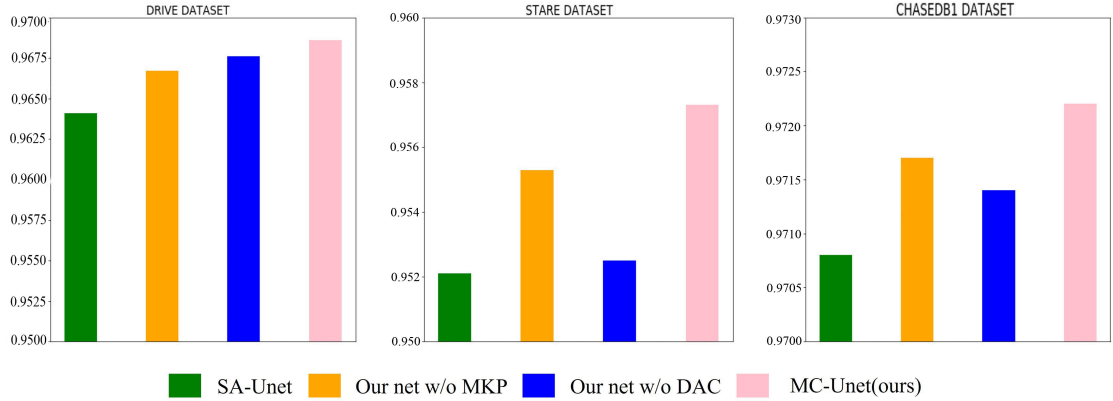

Fig 6. The accuracy indicator histogram. The Green bar represents the ACC results segmented by SA-Unet, the orange bar represents the ACC results by our net without MKP module, the blue bar represents the ACC results by our net without DAC module and the pink bar represents the ACC results segmented by MC-Unet in turn.

Table 3 shows the ablation experiments of the proposed model, comparing the backbone network (SA-UNet), SA-UNet + DAC, SA-UNet + MKP and MC-UNet. It it observed that the DAC module is able to enhance the specificity of the image effectively, reduce the blood vessels rate of false positive in the fundus image and the misdiagnosis cost of fundus image samples. The MKP module improves the AUC of the segmentation algorithm, making the algorithm more robust. Integrating the DAC and MKP modules into SA-UNet improves the segmentation effect as a whole, reduces the misdiagnosis rate of the image, and improves the ability to predict blood vessels by the algorithm. Figure 6 more intuitively shows the change of ACC in ablation experiment. Figure 7 compares the ROC curves of five different methods on three data sets. It can be seen from the results that our method achieves the best effect.

Table 3. The Ablation experiment results (%) of vessel segmentation on DRIVE, CHASEDB1 and STARE dataset.

| | | DRIVE | | | CHASEDB1 | | | STARE | | |
|---|---|---|---|---|---|---|---|---|---|---|
| DAC | MKP | ACC | SP | AUC | ACC | SP | AUC | ACC | SP | AUC |
| | | 96.41 | 97.67 | 97.38 | 97.08 | 98.09 | 97.78 | 95.21 | 99.30 | 96.26 |
| √ | | 96.67 | **98.90** | 97.96 | 97.17 | 98.22 | 98.17 | 95.28 | 98.93 | 95.83 |
| | √ | 96.72 | 98.29 | 98.42 | **97.22** | 98.10 | 98.29 | 95.58 | 99.23 | 96.50 |

| | | 96.78 | 98.79 | 98.28 | 97.14 | 98.29 | 98.18 | 95.72 | 99.47 | 96.86 |

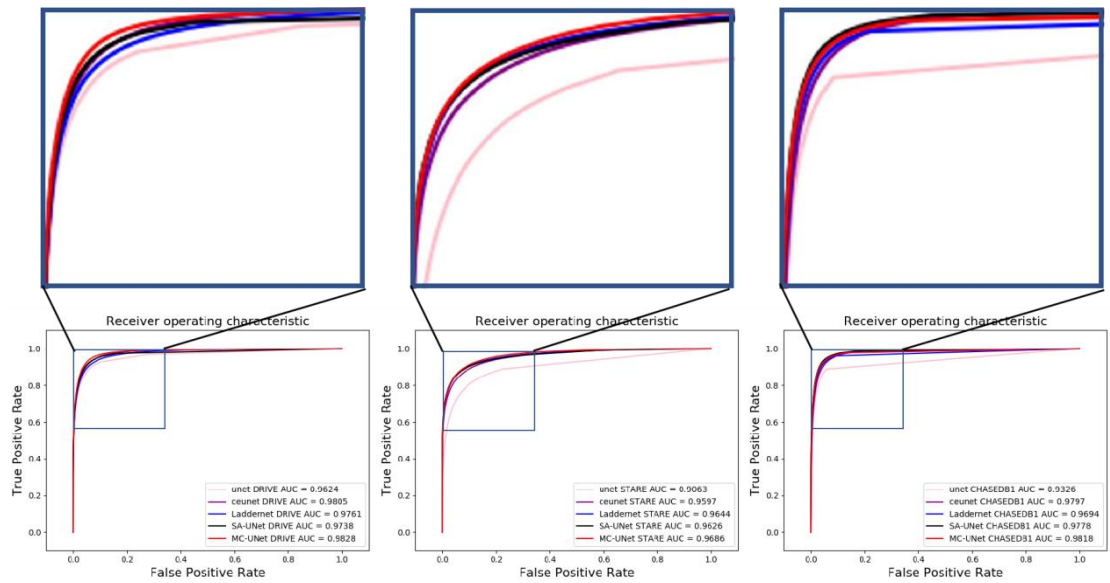

Fig 7. The ROC curves of different method on DRIVE (left), STARE (middle) and CHASEDB1 (right) datasets.

## 4. Conclusion

In order to solve the limited comprehensive features extracted by the encoder-decoder structure in the U-shaped network, which may lead to the segmentation loss of small fragile capillaries, a novel U-shape network is proposed named Multi-module Concatenation U-Net (MC-UNet) based on atrous convolution and multi-kernel pooling for retinal vessels segmentation. The network retains the local and global information about the main retinal vessels and capillaries. The DAC and MKP module are introduced to increase the receptive field for improving the sensitivity of the algorithm and retain more detailed feature information for improving the accuracy of retinal vascular segmentation. Experimental results proves the effectiveness of the method, especially for microvessels. However, For more severe lesions image data, a robust framework is still needed to be studied and discussed.

## 5. References

[1] Kim J H , Cho H K . The Influences of Arteriosclerosis on the Development and Progression of Diabetic Retinopathy[J]. 1999.

[2] https://www.who.int/news-room/fact-sheets/detail/blindness-and-visual-impairment

[3] Yu D Y , Yu P K , Cringle S J , et al. Functional and morphological characteristics of the retinal and choroidal vasculature.[J]. Progress in Retinal & Eye Research, 2014, 40:53-93.

[4] Chang-Ying L I , Chen N , Shi L P , et al. Morphological characteristics of anterior visual pathway and the relationship between morphological changes and retinal nerve fibre layer thickness in patients with primary open-angle glaucoma[J]. Chinese Journal of Medical Imaging Technology, 2010, 26(9):1628-1631.

[5] Saffarzadeh, Mohammadi V , Osareh, et al. Vessel Segmentation in Retinal Images Using Multi-scale Line Operator and K-Means Clustering.[J]. Journal of Medical Signals & Sensors, 2014.

[6] Wang Y B , Zhu C Z , Yan Q F , et al. A Novel Vessel Segmentation in Fundus Images Based on SVM[C]// International Conference on Information System & Artificial Intelligence. IEEE, 2017.

[7] Zhu C , Yao X , Zou B , et al. Retinal Vessel Segmentation in Fundus Images Using CART and AdaBoost[J]. Jisuanji Fuzhu Sheji Yu Tuxingxue Xuebao/Journal of Computer-Aided Design and Computer Graphics, 2014, 26(3):445-451.

[8] Oliveira W S , Teixeira J V , Ren T I , et al. Unsupervised Retinal Vessel Segmentation Using Combined Filters[J]. Plos One, 2016, 11(2):e0149943.

[9] Alhussein M , Aurangzeb K , Haider S I . An Unsupervised Retinal Vessel Segmentation Using Hessian and Intensity Based Approach[J]. IEEE Access, 2020, 8:165056-165070.

[10] Azzopardi G , Strisciuglio N , Vento M , et al. Trainable COSFIRE filters for vessel delineation with application to retinal images[J]. Medical Image Analysis, 2015, 19(1):46-57.

[11] Ronneberger, Olaf, Philipp Fischer and Thomas Brox. "U-Net: Convolutional Networks for Biomedical Image Segmentation." MICCAI (2015).

[12] Li L , Verma M , Nakashima Y , et al. IterNet: Retinal Image Segmentation Utilizing Structural Redundancy in Vessel Networks[C]// 2020 IEEE Winter Conference on Applications of Computer Vision (WACV). IEEE, 2020.

[13] Alom M Z , Yakopcic C , Hasan M , et al. Recurrent residual U-Net for medical image segmentation[J]. Journal of Medical Imaging, 2019, 6(1).

[14] Zhuang J . LadderNet: Multi-path networks based on U-Net for medical image segmentation[J]. arXiv e-prints, 2018.

[15] Wang B , Qiu S , He H . Dual Encoding U-Net for Retinal Vessel Segmentation[C]// International Conference on Medical Image Computing and Computer-Assisted Intervention. Springer, Cham, 2019.

[16] Guo C , Szemenyei M , Yi Y , et al. SA-UNet: Spatial Attention U-Net for Retinal Vessel Segmentation[J]. 2020.

[17] Gu Z , Cheng J , Fu H , et al. CE-Net: Context Encoder Network for 2D Medical Image Segmentation[J]. IEEE Transactions on Medical Imaging, 2019:1-1.

[18] Oktay O , Schlemper J , Folgoc L L , et al. Attention U-Net: Learning Where to Look for the Pancreas[J]. 2018

[19] Chen L C , Papandreou G , Kokkinos I , et al. Semantic Image Segmentation with Deep Convolutional Nets and Fully Connected CRFs[J]. Computer Science, 2014(4):357-361.

[20] Chen L C , Papandreou G , Kokkinos I , et al. DeepLab: Semantic Image Segmentation with Deep Convolutional Nets, Atrous Convolution, and Fully Connected CRFs[J]. IEEE Transactions on Pattern Analysis and Machine Intelligence, 2018, 40(4):834-848.

[21] Y. Yuan, L. Zhang, L. Wang and H. Huang, "Multi-Level Attention Network for Retinal Vessel Segmentation," in IEEE Journal of Biomedical and Health Informatics, vol. 26, no. 1, pp. 312-323, Jan. 2022, doi: 10.1109/JBHI.2021.3089201.

[22] Staal J, Abramoff MD, Niemeijer M et al. Ridge-based vessel segmentation in color images of the retina. IEEE Transactions on Medical Imaging 2004;23(4): 501–509. DOI:10.1109/TMI. 2004.825627.

[23] Hoover, Adam, Kouznetsova et al. Locating blood vessels in retinal images by piecewise threshold probing of a matched filter response. IEEE Transactions on Medical Imaging 2000.

[24] Owen, C. G. , et al. "Measuring retinal vessel tortuosity in 10-year-old children: validation of the Computer-Assisted Image Analysis of the Retina (CAIAR) program. " Investigative Ophthalmology & Visual Science 50.5(2009):2004-10.

[25] Kaiming, He, Xiangyu, et al. Spatial Pyramid Pooling in Deep Convolutional Networks for Visual Recognition[J]. IEEE Transactions on Pattern Analysis & Machine Intelligence, 2015.